\documentclass{desyprocA4}
\usepackage{heppennames2}

\begin{document}

\newcommand{\hbb}{\ensuremath{\PH\to b\bar{b}}\xspace}
\newcommand{\hcc}{\ensuremath{\PH\to c\bar{c}}\xspace}
\newcommand{\hmumu}{\ensuremath{\PH\to \mu^+\mu^-}\xspace}
\newcommand{\epemmumu}{\ensuremath{e^+e^-\to \mu^+\mu^-}\xspace}
\newcommand{\clicsid}{CLIC\_SiD\xspace}

\newcommand{\micron}{\ensuremath{\upmu\mathrm{m}}}
\newcommand{\degrees}{\ensuremath{^{\circ}}\xspace}

\newcommand{\mumu}{\ensuremath{\mu^+\mu^-}\xspace}
\newcommand{\mpmm}{\ensuremath{\PGmp\PGmm}\xspace}  
\newcommand{\tptm}{\ensuremath{\PGtp\PGtm}\xspace} 
\newcommand{\gamgam}{\ensuremath{\upgamma\upgamma}\xspace}
\newcommand{\gghadrons}{\ensuremath{\upgamma\upgamma \rightarrow \mathrm{hadrons}}\xspace}

\newcommand{\nuenuebar}{\ensuremath{\PGne\PAGne}\xspace} 
\newcommand{\nunubar}{\ensuremath{\PGn\PAGn}\xspace}   
\newcommand{\bb}{\ensuremath{\mathrm{\PQb}\mathrm{\PAQb}}\xspace}
\newcommand{\cc}{\ensuremath{\mathrm{\PQc}\mathrm{\PAQc}}\xspace}
\newcommand{\abinv}{\ensuremath{\mathrm{ab}^{-1}}\xspace}
\newcommand{\fbinv}{\ensuremath{\mathrm{fb}^{-1}}\xspace}
\newcommand{\pbinv}{\ensuremath{\mathrm{pb}^{-1}}\xspace}
\newcommand{\kT}{\ensuremath{k_{t}}\xspace}
\newcommand{\pT}{\ensuremath{p_\mathrm{T}}\xspace}
\newcommand{\qqbar}{\ensuremath{q\bar{q}}\xspace}
\newcommand{\epem}{\ensuremath{e^{+}e^{-}}\xspace}

\title{Light Higgs Studies for the CLIC CDR}

\author{{\slshape Christian Grefe$^1$, Tomas Lastovicka$^2$, Jan Strube$^1$}\\[1ex]
$^1$CERN, 1211 Geneve 23, Switzerland  \\
$^2$Institute of Physics, Academy of Sciences, 182 21 Prague 8, Czech Republic }


\contribID{xy}


\confID{5380}  
\desyproc{LC-REP-2012-005}
\doi  

\maketitle

LC-REP-2012-005

\begin{abstract}
The Higgs boson is the most anticipated discovery at the LHC, which can only partially explore its true nature.
Thus one of the most compelling arguments to build a future linear collider is to investigate properties of the Higgs boson, especially to test the predicted linear dependence of the branching ratios on the mass of the final state.
At a 3\,TeV CLIC machine the Higgs boson production cross section is relatively large and allows for a precision measurement of the Higgs branching ratio to pairs of b and c quarks, and even to muons. The cross section times branching ratio of the decays \hbb, \hcc and \hmumu can be measured with a statistical uncertainty of approximately 0.22\%,  3.2\% and 15\%, respectively.

\end{abstract}

The electroweak symmetry breaking mechanism, called Higgs mechanism, predicts a fundamental spin-0 particle, whose existence currently is being investigated at the LHC. The answer to the question about its existence is expected in 2013. The Standard Model predicted linear dependence of the branching ratios on the mass of the final state could be altered by non-Standard Model couplings.
The LHC can deliver only very limited measurements of the Higgs sector, but a detailed exploration is crucial for a deep understanding of its nature.

The compact linear collider (CLIC) is a proposed \epem collider with a maximum centre-of-mass energy \mbox{$\sqrt{s} = 3$\,TeV}, based on a two-beam acceleration scheme~\cite{CLICacceleratorCDR}.
In the following we present the analysis of the measurement of the branching ratios \hbb, \hcc~\cite{lcd:2011-036} and \hmumu~\cite{lcd:grefeHmumu2011} at such a machine. The studies are based on fully simulated and reconstructed samples in the \clicsid~\cite{lcd:grefemuennich2011} detector concept and take into account the relevant background processes as well as the main beam-related background.

\section{The \clicsid detector concept}

The \clicsid detector, used in the full simulation of samples, is based on the SiD detector concept~\cite{Aihara:2009ad} developed for the ILC project. It is designed for particle flow calorimetry using highly granular calorimeters and has been adapted~\cite{lcd:grefemuennich2011} to the specific requirements at CLIC.

A superconducting solenoid with an inner radius of 2.9\,m provides a central magnetic field of 5\,T. The calorimeters are placed inside of the coil and consist of a 30 layer tungsten-silicon electromagnetic calorimeter with $3.5\times3.5$\,{mm$^2$} segmentation, followed by a tungsten-scintillator hadronic calorimeter with 75 layers in the barrel region and a steel-scintillator hadronic calorimeter with 60 layers in the endcaps. The read-out cell size in the hadronic calorimeters is $30\times30$\,{mm$^2$}. The iron return yoke outside of the coil is instrumented with 9 double RPC layers with $30\times30$\,{mm$^2$} read-out cells for muon identification.

The silicon-only tracking system consists of 5 $20\times20$\,{$\micron^2$} pixel layers followed by 5 strip layers with a pitch of 25\,{\micron}, a read-out pitch of 50\,{\micron} and a length of 92\,{mm} in the barrel region. The tracking system in the endcap consists of 5 strip disks with similar pitch and a stereo angle of 12\degrees, complemented by 7 pixelated disks of $20\times20$\,{$\micron^2$} in the vertex and far-forward region at lower radii.

\section{Analysis framework and data samples}

The physical processes are generated with the Whizard~\cite{Kilian:2007gr,whizard2} event generator, with fragmentation and hadronization done by Pythia~\cite{Sjostrand2006}. The full simulation and reconstruction is performed in the software framework of the \clicsid detector concept.

The event simulation is performed using SLIC~\cite{slic}, a wrapper for \textsc{Geant4}~\cite{Allison:2006ve}, while the reconstruction is done by lcsim and PandoraPFA packages.
We assume a total accumulated luminosty of 2\,ab$^{-1}$, corresponding to 4\,years of data taking at the nominal machine parameters. Table~\ref{tab:samples} lists the physics processes that were taken into account in the analyses, together with their cross sections and the number of simulated events.

\begin{table}
 \centering
\begin{tabular}{p{6cm} r r l}
\hline \hline
Process & $\sigma$ [fb] & $N_{\mathrm{events}}$ & Short label \\
\hline
$\epem \to \PH \nuenuebar$; \hmumu    & 0.120  &   21000 & \hmumu    \\
$\epem \to \PH \nuenuebar$; \hbb &   285  &   45000 & \hbb  \\
$\epem \to \PH \nuenuebar$; \hcc &    15  &  130000 & \hcc \\
\hline
$\epem \to \mpmm \nunubar$                       & $132^{*}$          & 5000000 & $\mpmm \nunubar$    \\
$\epem \to \mpmm \epem$                          & $346^{*}$          & 1350000 & $\mpmm \epem$       \\
$\epem \to \mpmm$                                &  $12^{*}$ 	   &   10000 & $\mpmm$             \\
$\epem \to \tptm$                                & $250^{*}$         &  100000 & $\tptm$             \\
$\epem \to \tptm \nunubar$                       & $125^{*}$         &  100000 & $\tptm \nunubar$    \\
$\epem \to q\bar{q}$								     & 3100	&   96000 & $qq$				  \\
$\epem \to q\bar{q}\nunubar$						     & 1300	&  170000 & $qq\nu\nu$	      \\
$\epem \to q\bar{q}\epem$							     & 3300	&   90000 & $qq\epem$			  \\
$\epem \to q\bar{q} e\bar{\nu_{e}}$						     & 5300	&   91000 & $qq e \nu$          \\
$\gamgam \to \mpmm$ (generator level only)       & $20000^{*}$        & 1000000 & $\gamgam \to \mpmm$ \\
\hline \hline
\end{tabular}
 \caption{List of processes considered for this analysis with their respective cross section $\sigma$ and the number of simulated events $N_{\mathrm{events}}$. The cross section takes into account the CLIC luminosity spectrum and initial state radiation. Cross sections marked with * include a cut on the invariant mass of the muon pair at generator level to lie between 100 and 140 GeV.}
 \label{tab:samples}
\end{table}

The dominant Higgs boson production channel at 3\,TeV is the WW fusion channel $\epem \to H \nu \bar{\nu}$ with a cross section of $\sigma_{H\nu\bar{\nu}} = 420$\,fb. The main background for all channels is the Z boson production via WW fusion, which has similar kinematics as the signal processes.

Beamstrahlung effects on the luminosity spectrum as well as initial and final state radiation are taken into account.
For the default configuration of a 3\,TeV CLIC~\cite{CLICacceleratorCDR}, 3.2 \gghadrons events per bunch crossing are expected on average.
With a spacing of 0.5 ns between bunches, these necessarily pile up in the subdetectors, for which we assume integration times of 10 ns,
except for the barrel hadronic calorimeter, which has an integration time of 100\,ns. To approximate the CLIC beam structure and background conditions, the equivalent of 60 bunch crossing of \gghadrons events were mixed with every simulated event. In the \hmumu analysis, only the signal sample was mixed with events from \gghadrons background.

For the processes involving jets in the final state, fragmentation products of the hadronic systems are forced to two jets using the exclusive \kT algorithm of the FastJet package~\cite{FastJet:2010}, where the parameter $R$ is set to~0.7.
The LCFI vertexing package~\cite{LCFI} is used to identify jets according to their quark content as \PQb, \PQc and light quarks and computes the corresponding jet flavour tag values.

The event classification in \hbb and \hcc analyses is based on the open source Fast Artificial Neural Network (FANN) package~\cite{fann}. FANN was modified to account for event weights during the neural network training. The event classification in the \hmumu analysis is done using the boosted decision tree (BDT) classifier implemented in the TMVA package~\cite{TMVA:2010}.

\section{Measurement of \hbb and \hcc}

The measurement of the \hbb and \hcc decays requires resolution of secondary vertices from the primary vertex and is thus an important benchmark of the vertex tracking detector design.

\subsection{Jet flavour tagging}

The flavour identification package developed by the LCFI~\cite{LCFI} collaboration consists of a topological vertex finder ZVTOP, which reconstructs secondary interactions, and a multivariate classifier which combines several jet-related variables to tag bottom, charm, and light quark jets.
Displaced vertices are the most significant characteristic of b quark decays. A combination of several vertex-related variables, complemented by additional track-related variables, form an input for the tagging classifier. A detailed description of the variables and the procedure are given in~\cite{LCFI}.

\begin{figure}
    \centering
    \includegraphics[width=.49\linewidth]{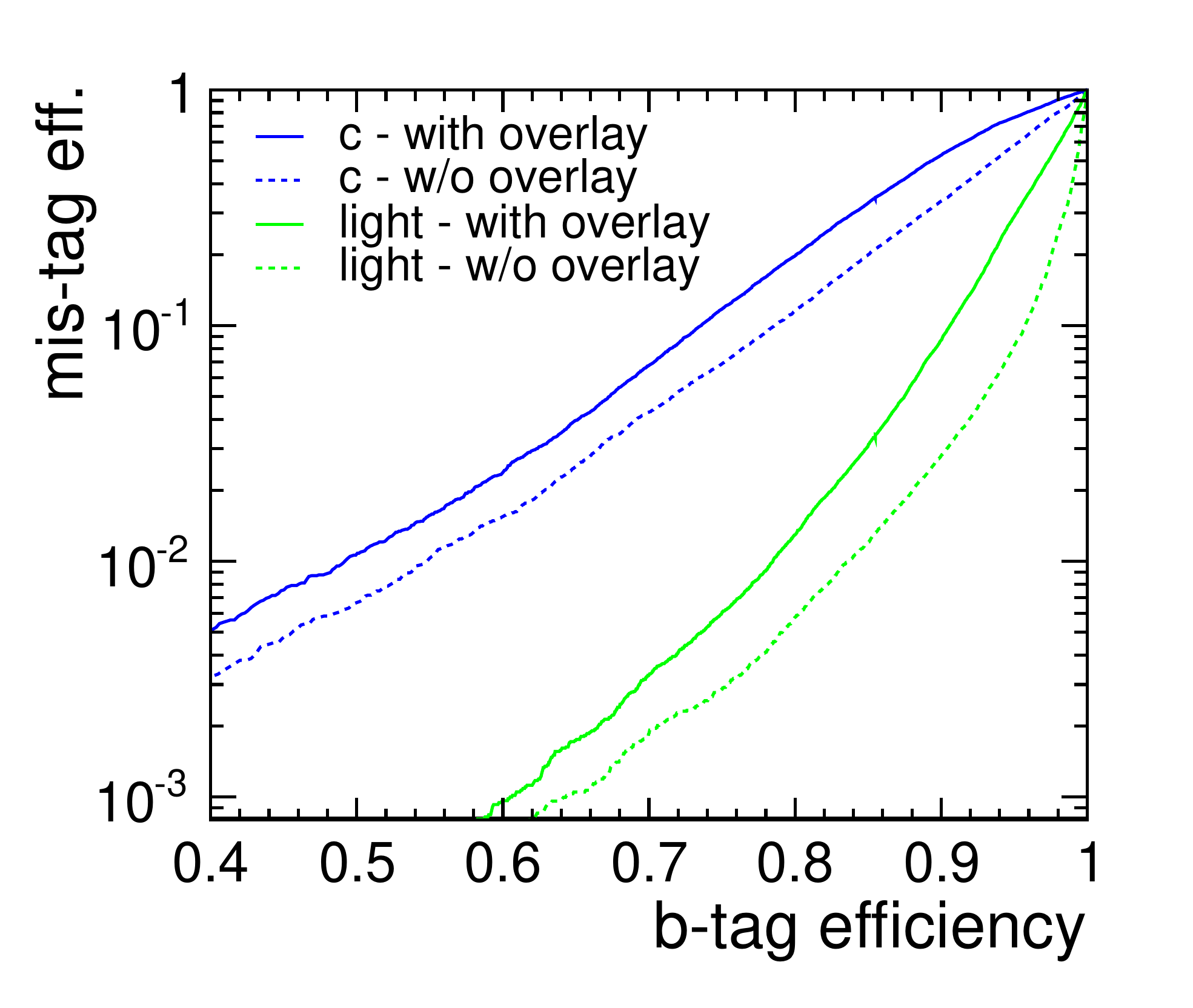} \hfill
    \includegraphics[width=.49\linewidth]{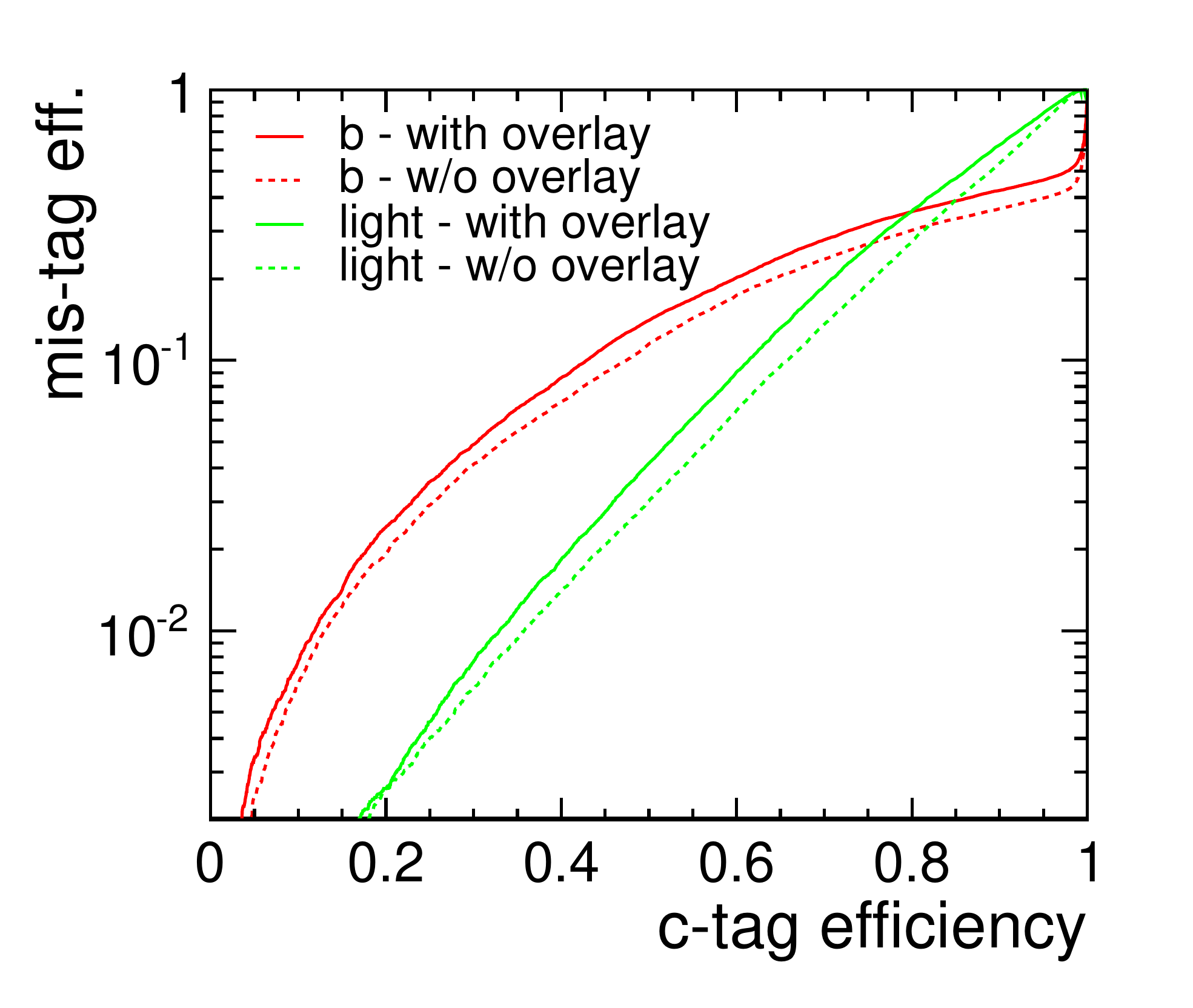}
    \caption{In the left plot, the mis-tag rate in the \clicsid detector for charm (blue) and light (green) jets as a function of the b-tag efficiency is shown. The right plot shows the mis-tag rate for bottom (red) and light (green) jets as a function of the c-tag efficiency. The mean \pT \,of the jets is 70\,{GeV} while the mean energy is $\sim130\,{GeV}$.}
    \label{fig:flavour_tag}
\end{figure}

The probability to tag a jet with a false flavour, the so called mis-tag rate, is used to assess the performance of the flavour tagging package. Figure~\ref{fig:flavour_tag} (left) shows the mis-tag rate for c-jets (blue line) and light jets (green line) as b-jets versus the b-tag efficiency, while Figure~\ref{fig:flavour_tag} (right) shows the mis-tag rate for b-jets (red line) and light jets (green line) as c-jets versus the c-tag efficiency. The presence of \gamgam backgrounds is found to reduce the flavour tagging performance, although the effect is not dramatic. The degradation of the flavour tag performance, shown in Figure~\ref{fig:flavour_tag}, has two sources: the flavour tag degradation itself plus a degradation of the jet quality due to a more difficult jet finding. For instance, at the b-tag efficiency of 70\% the mis-tag rate for c-jets (light jets) drops from 4.3\% (0.19\%) w/o overlay to 6.8\% (0.33\%) with overlay.

\subsection{Results}

The basic event selection requires two jets in each event. Apart from this selection, no further cuts are explicitly imposed and a number of relevant variables is given to a neural net for the subsequent multivariate analysis.
The invariant mass of the jet pair is the major discriminant between decays of Higgs and of Z bosons. It is used in a event classification neural network, together with the output of the b-flavor tagging network and the following variables:

\begin{itemize}\itemsep0pt
    \item The maximum of the absolute values of jet pseudorapidities.
    \item The sum of the remaining LCFI jet flavour tag values, i.e. c(udsb), c(b)-tags and b(uds)-tag\footnote{The notion indicates which flavour is tagged against which set of other flavours. For instance, c(b) is the c-flavour tagged against the b-flavour only, while remaining (uds) flavours are not used during the neural net training.}.
    \item $R_{\eta\phi}$, the distance of jets in the $\eta-\phi$ plane.
    \item The sum of jet energies.
    \item The total number of leptons in an event.
    \item The total number of photons in an event.
    \item Acoplanarity of jets.
\end{itemize}

Two neural nets were trained to separate either the \hbb or the \hcc signals from background samples accounting for event weights.
Thus the amount of the information about the signal, compared to the background, was proportional to its natural contribution. Such a solution delivers optimal results. It is more appropriate than, for instance, choosing the same number of signal and background events with no weights, or, training according to arbitrary sizes of the generated samples.

\begin{figure}
    \centering
    \includegraphics[width=.49\linewidth]{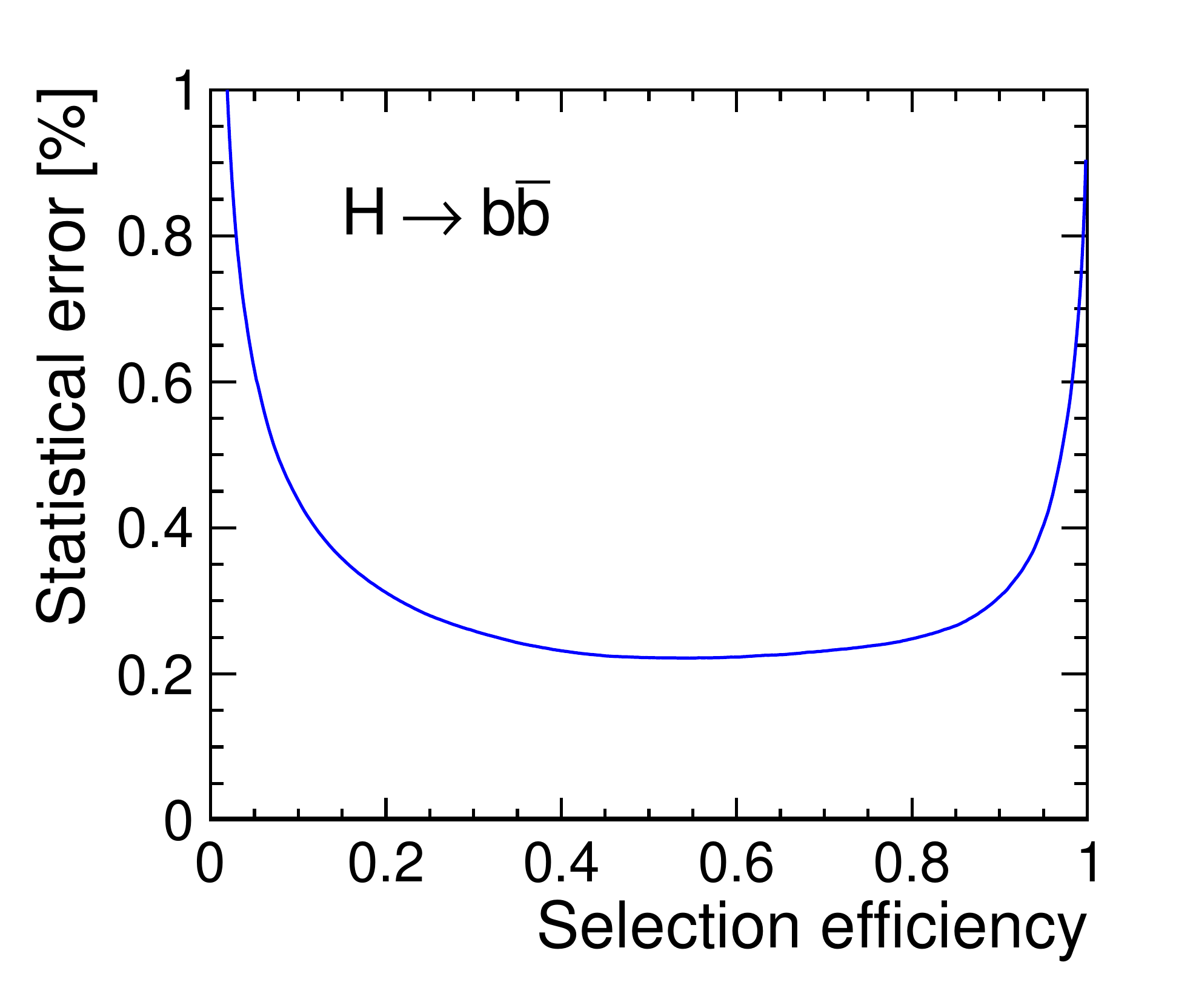}
    \includegraphics[width=.49\linewidth]{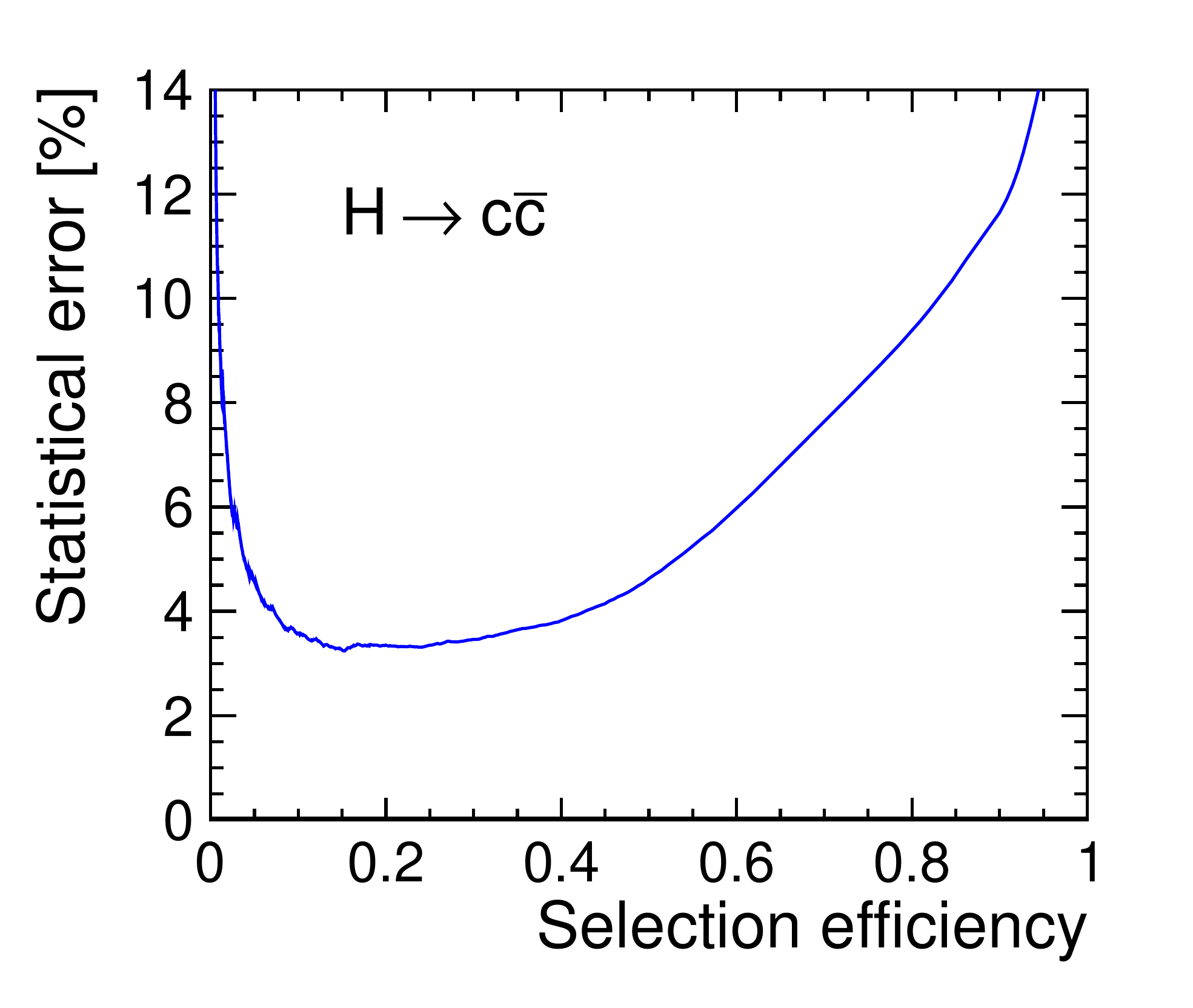}
    \caption{Statistical uncertainty of the measurement of cross section times branching ratio versus selection efficiency of the neural network. The neural network was trained to identify \hbb decays from di-jet backgrounds including \hcc (left). The neural network was trained on \hcc as signal and di-jets backgrounds including \hbb (right).}
    \label{fig:cross_section_error}
\end{figure}

The neural network selection efficiency versus the statistical uncertainty on the measurement is shown in Figure~\ref{fig:cross_section_error} for the two neural networks that were trained on \hbb and \hcc as signal, respectively. The optimal selection is at the local minimum of the curve, at a selection efficiency of 55\% for \hbb with a sample purity of 65\%, and a selection efficiency of 15\% for \hcc corresponding to a sample purity of 24\%. These values reflect the fact that b-jets can be distinguished from c-jets with high purity, while incompletely reconstructed b-jets and light jets make up a large fraction of the background to c-jet selection. Using the output of the reconstruction algorithms in neural networks leads to the minimal statistical uncertainty on the measurement at the eventual cost of an increased dependence on systematic effects. We assume that with sufficient experience at the running machine, the systematic variations are well enough understood so that the systematic uncertainties are comparable to the statistical uncertainties of the \hcc channel and dominate in the \hbb channel.

The resulting statistical \hbb cross section uncertainty amounts to $0.22\,\%$ while preserving meaningful values of both the sample purity (65.4\,\%) and of the signal selection efficiency (54.6\,\%).
The \hcc channel is more difficult to separate from the background and the statistical cross section uncertainty is $3.24\,\%$ with a signal selection efficiency of 15.2\%.

\section{Measurement of \hmumu}

The measurement of the rare decay \hmumu requires high luminosity operation and sets stringent limits on the momentum resolution of the tracking detectors. The branching ratio of the decay of a Standard Model Higgs boson to a pair of muons is important as the lower end of the accessible decays and defines the endpoint of the test of the predicted linear dependence of the branching ratios to the mass of the final state particles.

\begin{figure}
    \centering
    \includegraphics[width=.49\linewidth]{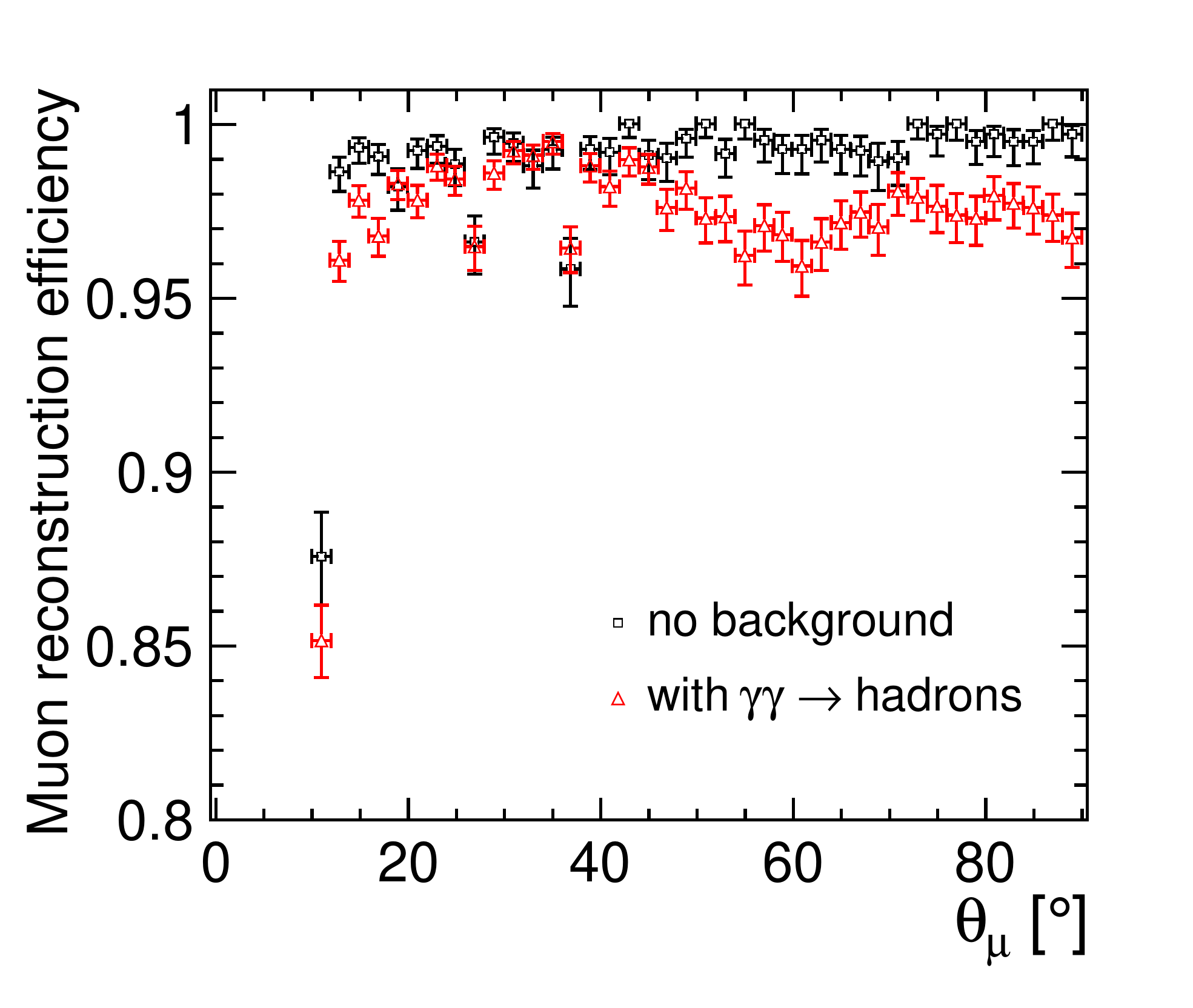}
    \includegraphics[width=.49\linewidth]{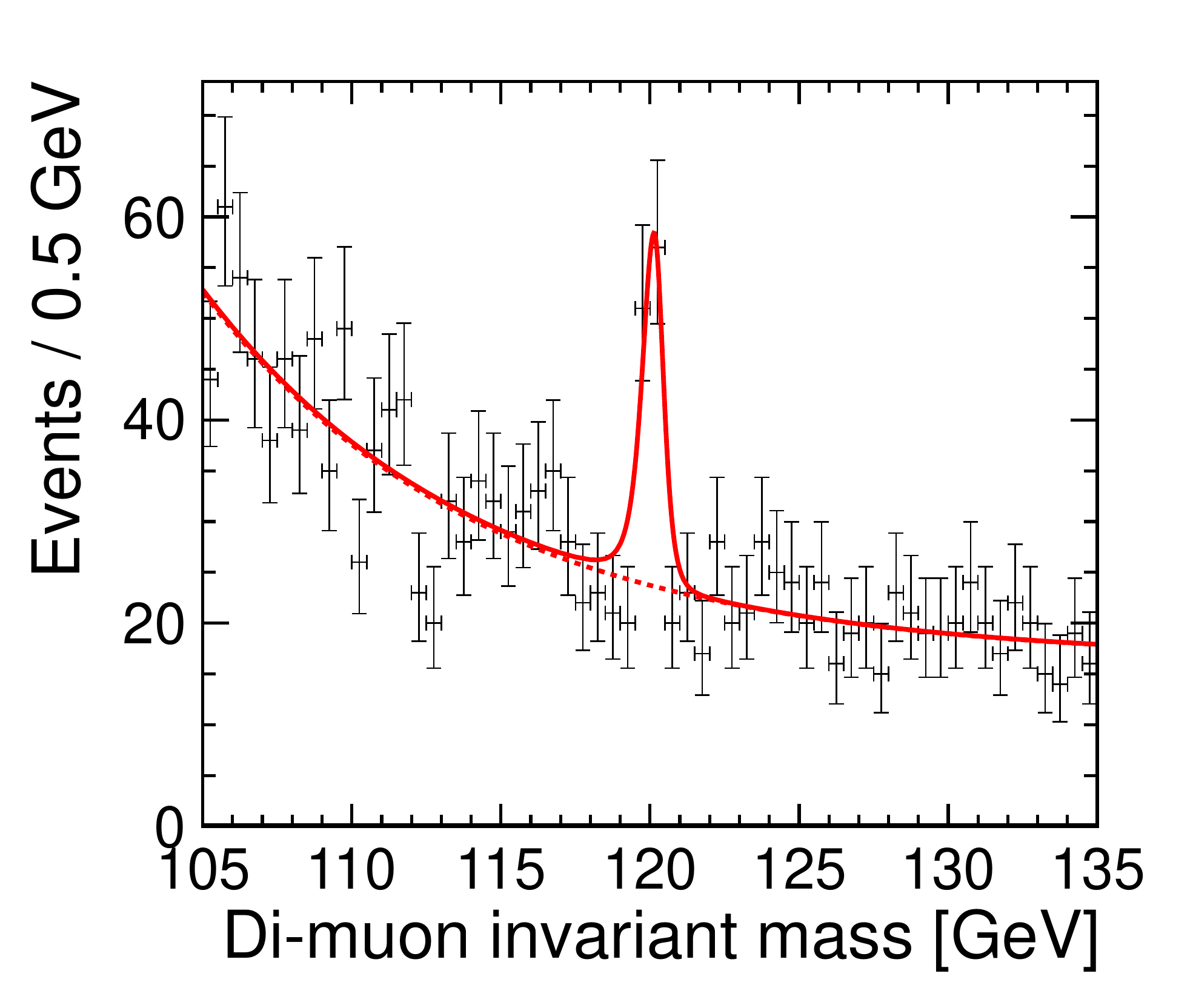}
    \caption{Muon reconstruction efficiency for the signal sample with and without \gghadrons pile-up.}
    \label{fig:muon_analysis}
\end{figure}

The events are selected by requiring two reconstructed muons, each with a transverse momentum of at least 5\,GeV.
In this note, the most energetic muon is referred to as $\upmu_1$ and the second most energetic muon is referred to as $\upmu_2$. In addition, the invariant mass of the two muons $M(\mumu)$ is required to lie between 105\,GeV and 135\,GeV.

The muon reconstruction efficiency is shown in Figure~\ref{fig:muon_analysis} (left). The beam induced background from \gghadrons leads to a small deterioration of the muon reconstruction efficiency. The average muon reconstruction efficiency for polar angles greater than 10\degrees is 98.4\% with this background compared to 99.6\% without. The total reconstruction efficiency of the signal sample, requiring two reconstructed muons with an invariant mass between 105\,GeV and 135\,GeV is 72\% in the presence of background.

The event classification is done using boosted decision tree classifier implemented in TMVA~\cite{TMVA:2010}.
The BDT is trained to separate the $\mpmm \nunubar$ signal events from the $\mpmm \epem$ background.
The $\mpmm$, $\tptm$ and $\tptm \nunubar$ samples are not used in the training of the BDT, but are effectively removed by the classifier nevertheless.

The variables used for the event selection by the BDT are:
\begin{itemize}\itemsep0pt
 \item The visible energy excluding the two reconstructed muons $E_{\mathrm{vis}}$.
 \item The scalar sum of the transverse momenta of the two muons $\pT(\upmu_1) + \pT(\upmu_2)$.
 \item The helicity angle $\cos\theta^*(\mumu) = \frac{\vec{p}'(\upmu_1) \cdot \vec{p}(\mumu)}{|\vec{p}'(\upmu_1)| \cdot |\vec{p}(\mumu)|}$, where $\vec{p}'$ is the momentum in the rest frame of the di-muon system. Since the two muons are back-to-back in the rest frame of the di-muon system there is no additional information to be gained from calculating a similar angle for $\upmu_2$.
 \item The relativistic velocity of the di-muon system $\upbeta(\mumu)$, where $\upbeta = \frac{v}{c}$.
 \item The transverse momentum of the di-muon system $\pT(\mumu)$.
 \item The polar angle of the di-muon system $\theta(\mumu)$.
\end{itemize}

The major discriminant is the visible energy whenever there is an electron within the detector acceptance. Otherwise the background can be rejected by the transverse momentum of the di-muon system or the sum of the two individual transverse momenta.
Figure~\ref{fig:muon_analysis} (right) shows the Higgs peak in the invariant mass distribution after the event selection.

The dominant background from $\epem \to \epem\mumu$ events, is effectively reduced by forward electron tagging. While the forward calorimeters were not part of the full detector simulation, assuming a tagging efficiency of 95\% down to an angle of 40 mrad for electrons of several hundred GeV to over one TeV is a conservative estimate, even in the presence of \gghadrons background. It is found that Bhabha events prevent further rejection of this background at lower angles. The results quoted are based on a ad-hoc rejection of 95\% of the electrons in the Luminosity Calorimeter.

\subsection{Invariant mass fit}
\label{sec:MassFit}	
The distribution of the invariant mass in the \hmumu sample has a tail towards lower masses because of final state radiation. The shape can be described best by two half Gaussian distributions with an exponential tail. Together with the mean value this results in five free parameters in the fitted function, which can be written as
\begin{equation*}
 f(x) = n \left\{ \begin{array}{rl}
                 e^{\frac{-(x - m_0)^2}{2\sigma_L^2 + \alpha_L(x - m_0)^2}} &\mbox{, $x \leq m_0$} \\
                 e^{\frac{-(x - m_0)^2}{2\sigma_R^2 + \alpha_R(x - m_0)^2}} &\mbox{, $x > m_0$}
                \end{array} \right . ,
 \label{eq:MassFit_hmumu}
\end{equation*}
where $m_0$ is the mean of both Gaussian distributions, $\sigma_L$ and $\sigma_R$ are the widths, and $\alpha_L$ and $\alpha_R$ are the tail parameters of the left and the right Gaussian distribution, respectively; $n$ is a normalization parameter.
The background is well described by an exponential parameterization, obtained from a background-only sample.

The number of signal events is obtained from a maximum likelihood fit to the sample containing signal plus background after the event selection.

The average muon momentum resolution of the fully simulated sample is $4 \times 10^5$\,GeV$^{-1}$ corresponding to a statistical uncertainty of
23\% without the forward electron tagging. If the background from $\epem\to\mpmm$ can be reduced using tagging of electrons down to an angle of 40 mrad with an efficiency of 95\%, the cross section times branching ratio can be measured to a precision of 15\%.

\section{Summary}

The sensitivity to the decay branching ratios of a neutral 120\,GeV Standard Model Higgs boson to bottom and charm quarks and to muons has been studied at the CLIC centre-of-mass energy of \mbox{$\sqrt{s} = 3$\,TeV} and integrated luminosity of 2\,ab$^{-1}$. The analysis is based on full simulation and realistic event reconstruction in the \clicsid detector. We have demonstrated the feasibility of such measurements and estimated their statistical uncertainty.

For the measurement of Higgs decays to quarks, 0.22\% and 3.2\% statistical uncertainty can be achieved for the decays \hbb and \hcc, respectively. This includes the effect of background from \gghadrons on the flavor tagging.

For the rare decay \hmumu, the cross section times branching ratio can be measured to a precision of 15\% if the background from \epemmumu can be reduced using tagging of electrons down to an angle of 40 mrad with an efficiency of 95\%.
The effect of \gghadrons has been taken into account conservatively by only including it in the signal sample and thus reducing its reconstruction efficiency.

From experience of the LEP experiments one can assume that the systematic uncertainties related to detector effects are of the order of 1\% or less. For the measurement of $\sigma_{\ensuremath{\PZz\to \mpmm}}$ at LEP the systematic uncertainty was between 0.1 and 0.4\%, depending on the experiment. Thus we expect that the systematic uncertainty of the \hmumu analysis will be negligible compared to the statistical uncertainty, the uncertainty of \hbb analysis will be dominated by the systematic and theoretical uncertainties and in the \hcc analysis the uncertainty sources will contribute comparably.


\begin{footnotesize}

\bibliographystyle{unsrt}
\bibliography{LightHiggs_CLIC}

\begin{thebibliography}{10}

\bibitem{CLICacceleratorCDR}
{The CLIC Accelerator Design, Conceptual Design Report;
  https://edms.cern.ch/document/1180032/}.

\bibitem{lcd:2011-036}
T.~Lastovicka.
\newblock {Light Higgs boson production and hadronic decays at 3 TeV}, 2011.
\newblock {CERN {LCD-Note-2011-036} }.

\bibitem{lcd:grefeHmumu2011}
C.~Grefe.
\newblock {Light Higgs decay into muons in the CLIC\_SiD CDR detector}, 2011.
\newblock {CERN {LCD-Note-2011-035}}.

\bibitem{lcd:grefemuennich2011}
C.~Grefe and A.~M\"unnich.
\newblock {The CLIC\_SiD\_CDR geometry for the CDR Monte Carlo mass
  production}, 2011.
\newblock {CERN {LCD-Note-2011-009} }.

\bibitem{Aihara:2009ad}
H.~Aihara et~al.
\newblock {SiD Letter of Intent}, 2009.
\newblock {{arXiv:0911.0006}, SLAC-R-944}.

\bibitem{Kilian:2007gr}
W.~Kilian, T.~Ohl, and J.~Reuter.
\newblock {WHIZARD: Simulating multi-particle processes at LHC and ILC}, 2007.
\newblock {arXiv:0708.4233v1 }.

\bibitem{whizard2}
M.~Moretti, T.~Ohl, and J.~Reuter.
\newblock {O'Mega: An optimizing matrix element generator}, 2001.
\newblock {arXiv:hep-ph/0102195v1}.

\bibitem{Sjostrand2006}
T.~Sjostrand, S.~Mrenna, and P.~Z. Skands.
\newblock {PYTHIA 6.4 Physics and Manual}.
\newblock {\em JHEP}, 05:026, 2006.
\newblock {hep-ph/0603175}.

\bibitem{slic}
Simulator for the Linear Collider (SLIC),
  {http://www.lcsim.org/software/slic/}.

\bibitem{Allison:2006ve}
J.~Allison et~al.
\newblock {Geant4 developments and applications}.
\newblock {\em IEEE Trans. Nucl. Sci.}, 53:270, 2006.

\bibitem{FastJet:2010}
M.~Cacciari and G.~P. Salam.
\newblock {Dispelling the $N^{3}$ myth for the $k_t$ jet-finder}.
\newblock {\em Phys. Lett.}, B641:57--61, 2006.
\newblock {hep-ph/0512210}.

\bibitem{LCFI}
A.~Bailey and {\it et al.}~(LCFI~Collaboration).
\newblock {LCFIVertex package: Vertexing, flavour tagging and vertex charge
  reconstruction with an ILC vertex detector}.
\newblock {\em Nucl. Instrum. Methods Phys. Res. A}, A 610:573--589, 2009.

\bibitem{fann}
Fast {A}rtificial {N}eural {N}etwork {L}ibrary ({FANN}).
\newblock {http://leenissen.dk/fann/wp/}.

\bibitem{TMVA:2010}
A.~H\"ocker, P.~Speckmayer, J.~Stelzer, J.~Therhaag, E.~von Toerne, and
  H.~Voss.
\newblock {TMVA - Toolkit for multivariate data analysis}, 2009.
\newblock {arXiv:physics/0703039}.

\end{thebibliography}

\end{footnotesize}

\end{document}